%% file: main.tex
\documentclass[acmlarge]{acmart} 

\AtBeginDocument{%
  }

\setcopyright{acmcopyright}
\copyrightyear{2025}
\acmYear{2025}

\acmDOI{XXXXXXX.XXXXXXX}

\acmVolume{1}
\acmNumber{1}
\acmMonth{1}

\usepackage{tikz}
\usepackage[utf8]{inputenc}
\usepackage[english]{babel}
\usepackage{booktabs} 
\usepackage[absolute,overlay]{textpos}
\usepackage{etoolbox}
\usepackage[figurename=Fig.,font={small},labelfont={bf,up}]{caption}
\usepackage[export]{adjustbox}
\usepackage[linesnumbered,ruled,vlined]{algorithm2e}
\usepackage{color, colortbl}
\usepackage{longtable}
\usepackage{xcolor}
\usepackage{soul}
\usepackage{comment}
\usepackage{url}
\usepackage{balance}
\usepackage{graphicx}
\usepackage{subfigure}
\usepackage{multirow}
\usepackage{textcomp}
\usepackage{lipsum}
\usepackage{tikz}
\usepackage{hyperref}
\usepackage{cleveref}
\usepackage{float}
\usepackage{xspace}
\usepackage{stfloats}
\usepackage{enumitem}
\usepackage{arydshln}

\newbool{showComments}
\booltrue{showComments}

\ifbool{showComments}{%

}

\newcommand{\SysName}{CP-PPG\xspace}

\begin{document}

\title{Reliable Physiological Monitoring on the Wrist Using
Generative Deep Learning to Address Poor
Skin-Sensor Contact}

\author{Manh Pham Hung}
\authornote{Both authors contributed equally to this research.}
\email{hm.pham.2023@phdcs.smu.edu.sg}
\affiliation{%
  \institution{Singapore Management University}
  \country{Singapore}
}

\author{Matthew Yiwen Ho}
\authornotemark[1]
\email{matthewho.2021@scis.smu.edu.sg}
\affiliation{%
  \institution{Singapore Management University}
  \country{Singapore}
}

\author{Yiming Zhang}
\affiliation{%
  \institution{The Chinese University of Hong Kong}
  \country{China}
}

\author{Dimitris Spathis}
\affiliation{%
 \institution{University of Cambridge}
 \country{United Kingdom}}

\author{Aaqib Saeed}
\affiliation{%
  \institution{Eindhoven University of Technology}
  \country{Netherlands}}

\author{Dong Ma}
\orcid{}
\authornote{Corresponding author: Dong Ma, dongma@smu.edu.sg, Singapore Management University}
\affiliation{%
  \institution{Singapore Management University}
  \country{Singapore}}
\email{dongma@smu.edu.sg}

\input{sections/0-Abstract}



\maketitle

\input{sections/01-Introduction}

\input{sections/02-Preliminary}

\input{sections/03-SystemDesign}

\input{sections/06-PerformanceEvaluation}

\input{sections/08-Discussion}

\input{sections/07-RelatedWorks}

\input{sections/09-Conclusion}

\bibliographystyle{ACM-Reference-Format}
\bibliography{main}

\appendix

\end{document}

%% file: sections/0-Abstract.tex
\begin{abstract}
Photoplethysmography (PPG) is a widely adopted, non-invasive technique for monitoring cardiovascular health and physiological parameters in both consumer and clinical settings. While motion artifacts in dynamic environments have been extensively studied, suboptimal skin–sensor contact in sedentary conditions - a critical yet underexplored issue - can distort PPG waveform morphology, leading to the loss or misalignment of key features and compromising sensing accuracy. In this work, we propose \SysName, a novel framework that transforms \underline{C}ontact \underline{P}ressure–distorted \underline{PPG} signals into high-fidelity waveforms with ideal morphology. \SysName integrates a custom data collection protocol, a carefully designed signal processing pipeline, and a novel deep adversarial model trained with a custom PPG-aware loss function. We validated \SysName through comprehensive evaluations, including 1) morphology transformation performance on our self-collected dataset, 2) downstream physiological monitoring performance on public datasets, and 3) in-the-wild study. Extensive experiments demonstrate substantial and consistent improvements in signal fidelity (Mean Absolute Error: 0.09, 40\% improvement over the original signal) as well as downstream performance across all evaluations in Heart Rate (HR), Heart Rate Variability (HRV), Respiration Rate (RR), and Blood Pressure (BP) estimation (on average, 21\% improvement in HR; 41-46\% in HRV; 6\% in RR; and 4-5\% in BP). These findings highlight the critical importance of addressing skin-sensor contact issues to enhance the reliability and effectiveness of PPG-based physiological monitoring. \SysName thus holds significant potential to improve the accuracy of wearable health technologies in clinical and consumer applications.

\end{abstract}

%% file: sections/01-Introduction.tex
\section{Introduction} \label{introduction}
\begin{figure}[h]
    \centering    
    \includegraphics[width=0.54\linewidth]{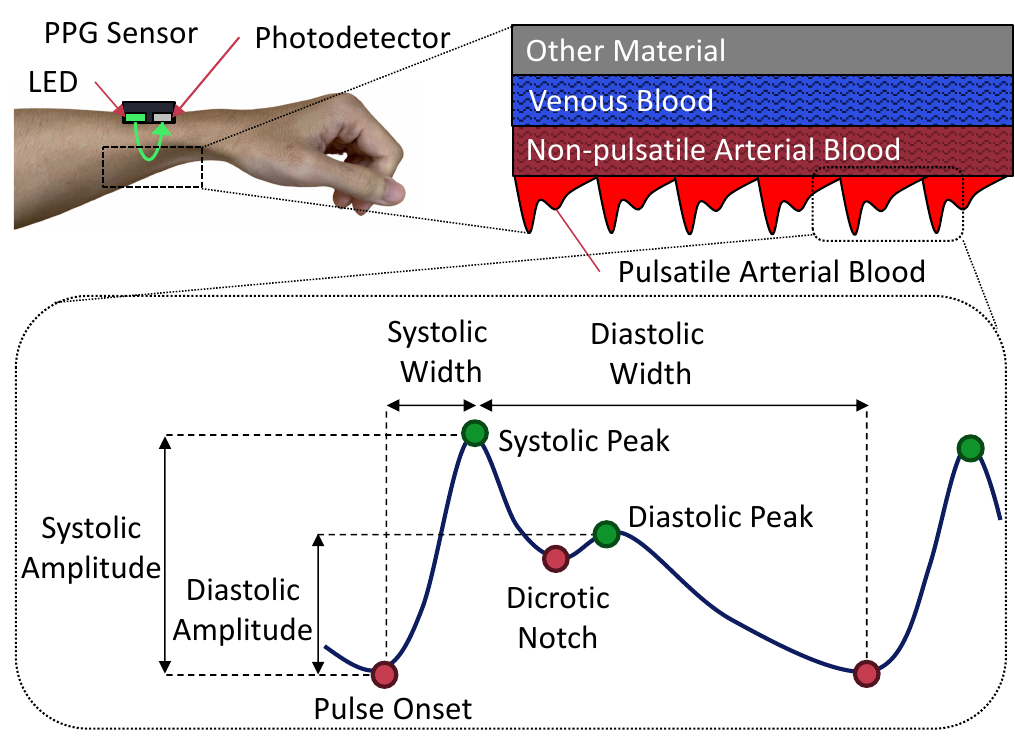}
    \caption{PPG principle and the ideal PPG morphology with the presence of systolic peak, diastolic peak, and dicrotic notch.}
    \label{fig:PPG_Principle_and_Waveform}
\end{figure}

Photoplethysmography (PPG) is a non-invasive optical technique used to measure changes in blood volume in peripheral blood vessels. It functions by emitting light onto the skin using an LED and measuring the amount of light transmitted through (transmissive PPG) or reflected against (reflective PPG) the skin, which changes due to cardiac pulsations, with a photodetector. An \textbf{ideal PPG waveform} (\Cref{fig:PPG_Principle_and_Waveform}) reflects complete cardiac cycles and contains all key features such as the systolic peak, dicrotic notch, and diastolic peak, all of which hold significant clinical importance. For example, PPG waveforms can be used to derive various physiological parameters, including heart rate (HR)~\cite{8703846, 8571310, arunkumar2019heart}, heart rate variability (HRV)~\cite{natarajan2020heart, taoum2022validity, hoog2021accuracy, aliani2023automatic}, respiration rate (RR)~\cite{osathitporn2023rrwavenet, natarajan2021measurement}, blood pressure (BP)~\cite{10.1145/3610918,finnegan2023features, elgendi2019use,joung2023continuous}, and blood oxygen saturation (SpO2)~\cite{chu2023non}, to detect infectious diseases~\cite{karolcik2024towards}, as well as to provide valuable information about cardiovascular functions such as artery stiffness~\cite{Shin2022ArterialStiffness, ferizoli2024arterial}, atrial fibrillation~\cite{10.1145/3517247}, diabetes~\cite{Gupta2022Diabetes}, and chronic obstructive pulmonary disease (COPD)~\cite{Davies_2022_COPD}. Furthermore, PPG signals have been shown to potentially enable applications such as biometric authentication~\cite{10.1145/3161189} and facial expression recognition~\cite{10.1145/3534597}.

With the advancement of mobile health, PPG has emerged as a leading technology for continuous health monitoring on wearable devices. For instance, many off-the-shelf smartwatches (e.g., Apple Watch \cite{applewatch}, Samsung Galaxy Watch \cite{samsungwatch}, and others) and even some recent earphones~\cite{ppgEarphones} are equipped with PPG sensors for estimating HR, HRV, BP, and SpO2. Despite its successful adoption in commercial products, PPG remains an active research area due to its susceptibility to various real-world factors that can compromise its sensing performance. A primary focus of prior research has been combating motion artifacts, defined as PPG signal distortions caused by the movement of users during physical activities like walking and running~\cite{parak2017estimating, temko2017accurate, stahl2016accurate, he2019robust}. Proposed solutions, such as adaptive filtering, accelerometer-based signal compensation, and deep learning algorithms, have shown promising results~\cite{Peng2014AdaptiveFilter, Pollreisz2019AccelerometerCompensation, afandizadeh2023accurate, zheng2024tiny}. 

Nevertheless, the sensing performance of PPG in \textbf{sedentary} scenarios has often been overlooked. For instance, as shown in \Cref{fig:5Poses}, in smartwatch-based PPG sensing, individuals may adopt various wrist postures during different sedentary activities (e.g., reading), resulting in changes in contact pressure (CP) between the PPG sensor and human skin (i.e., tightness). 
Previous studies have shown that variations in CP can affect the accuracy of physiological parameter estimation through field experiments~\cite{Chandrasekhar2020ContactPressure, Grabovskis2013PressureOnWaveforms}, highlighting the importance of considering CP during estimation. \textit{However, these studies have primarily focused on the influence of CP on signal amplitude (or signal-to-noise ratio) while 1) ignoring its impact on other key waveform features (e.g, presence of dicrotic notch, systolic/diastolic width as shown in~\Cref{fig:PPG_Principle_and_Waveform}) and 2) without proposing any solutions to mitigate these impacts.} 

In our study, we observe that, in addition to signal amplitude, the \textbf{entire morphology} of the PPG waveform can be influenced by CP. Although these distortions are less extreme than those caused by motion, they prevent the extraction of more subtle yet important features for complex PPG applications. For example, as depicted in \Cref{fig:5Poses}, certain wrist postures (e.g., hand on mouse) can cause temporal shifts in peak positions, which negatively impact HR and HRV estimation that depends on precise timing. Meanwhile, in some postures (e.g., reading a book and holding a pen), the dicrotic notch and diastolic peak may be diminished or even absent altogether. Consequently, downstream applications relying on these features (e.g., BP estimation) may be rendered unfeasible even in seemingly ideal, motionless conditions.

Therefore, developing a robust solution to enhance the quality of PPG signals under suboptimal CP is essential. To address this, we propose a deep learning-based framework called \SysName, designed to restore the ideal PPG morphology, typically observed under optimal CP, from distorted signals measured under sedentary conditions with suboptimal CP. In designing \SysName, we faced two practical challenges: 1) training such a deep neural network (DNN) requires simultaneously recorded CP-distorted and ideal signals, which are unavailable in public PPG datasets and impossible to collect from the same position simultaneously; 2) effectively recovering missing features (e.g., diastolic peak and dicrotic notch) is complex and demands a careful modeling design. 

\SysName addresses these challenges by: 1) developing a customized wearable prototype for data collection from 22 subjects. The collected data includes two synchronously measured PPG streams - one from the wrist and the other from the finger. The wrist PPG was exposed to various contact pressures to produce a diverse range of PPG morphologies, while the finger PPG was maintained at an optimal CP to obtain the ideal morphology; 2) devising a novel autoencoder-based DNN and training it with a custom loss function incorporating PPG domain knowledge and adversarial training techniques. 
We evaluated \SysName through comprehensive experiments and demonstrated its superior performance in transforming distorted morphology into 
forms more accurately representing physiological attributes on the collected data. Additionally, with experiments on four public PPG datasets and a longitudinal in-the-wild study, we showed that the transformed waveforms enable more accurate estimation of typical physiological metrics, including HR, HRV, BP, and RR. These experiments also highlight the generalizability of \SysName, positioning it as a valuable \textbf{plug-in API} for refining PPG morphology for more accurate and robust PPG sensing in unseen
conditions.

We summarize the contributions of this work as follows:
\begin{itemize}[left=0.2cm]
  \item To the best of our knowledge, we are the first to address the impact of contact pressure on PPG signal morphology and downstream physiological sensing applications.
  
  \item We developed a comprehensive data pipeline and a novel DNN structure that particularly leverages PPG domain knowledge for effective PPG morphology transformation. 
  
  \item  With extensive experiments on our self-collected data, public datasets, and in-the-wild longitudinal study, we demonstrated the effectiveness and generalizability of the proposed approach. 
\end{itemize}

%% file: sections/02-Preliminary.tex
\section{Preliminary} \label{preliminary}

\begin{figure}[h]
    \centering    
    \includegraphics[width=0.64 \linewidth]{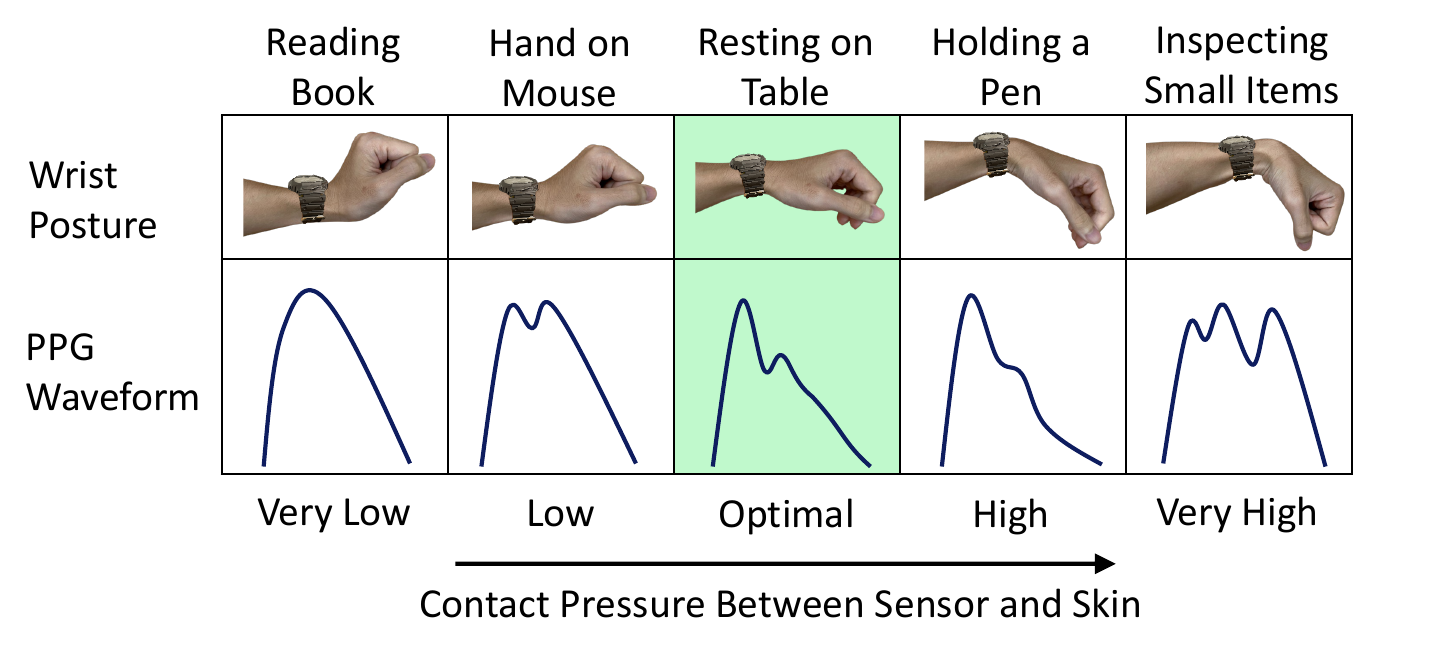}
    \caption{Effects of wrist postures on CP and PPG morphology.}
    \label{fig:5Poses}
\end{figure}

\subsection{PPG Morphology vs. Downstream Tasks}
\label{sec: pre_downstream}

PPG signals encompass both pulsatile and non-pulsatile components, also known as the alternating current (AC) and direct current (DC) components, respectively~\cite{Kao_Chao_Wey_2019}. The AC component reflects blood volume variations attributable to heartbeats, while the DC component comprises the sum of light reflected from the baseline blood volume between heartbeats and non-pulsatile materials. As illustrated in~\Cref{fig:PPG_Principle_and_Waveform}, the \textbf{ideal morphology} of a PPG waveform (AC component) contains three key points: the \textbf{systolic peak} (representing the peak pressure generated by the contraction of the heart's left ventricle during systole), the \textbf{dicrotic notch} (resulting from the closure of the aortic valve), and the \textbf{diastolic peak} (during the diastole phase when the heart muscle relaxes). These points' amplitudes and positions constitute a variety of temporal and spatial features such as pulse width, peak-to-peak distance, diastolic peak amplitude, and etc, which hold significant clinical importance in assessing cardiac and vascular health~\cite{Chowdhury2020bpWithML}. 

Specifically, researchers have showcased numerous downstream tasks that can be realized with features extracted from the PPG waveform. For example, HR can be derived by counting the number of systolic peaks or waveform feet using peak-detection algorithms within a specific timeframe \cite{Yu2006PPG-HR}. Pulse widths, or the time between waveform beginnings, can also be used in HRV calculation algorithms \cite{Jeyhani2015PPGHRV}. Rhythmic variations in pulse widths have also been used to estimate respiratory rate \cite{Iqbal2022RR}. 
Recently, researchers have extracted various hand-crafted features from the PPG waveform (e.g., peak positions, distance between peaks, waveform widths, and waveform slope characteristics) and leveraged machine learning algorithms to estimate blood pressure \cite{Chowdhury2020bpWithML} and blood glucose \cite{Hammour2023Glucose}. Moreover, the second derivative of the PPG signal can be used to estimate vascular ageing~\cite{Rozi2012SDPPG-Age}.

\subsection{PPG Morphology vs. Contact Pressure}
Although the ideal morphology of the PPG waveform allows for the extraction of various fine-grained features, it is often absent in real-world scenarios due to the susceptibility of the PPG sensor to adverse measurement conditions. Specifically, a substantial body of prior research has demonstrated that PPG signals are often corrupted or completely overshadowed during intensive physical activities (e.g. running)~\cite{Seok2021PPGMA, Kong2019, Boloursaz2016, Shimazaki2014}. More recently, researchers have noticed that the contact pressure between the PPG sensor and human skin also impacts the PPG waveform, which causes the loss of critical waveform features as mentioned above (e.g., the amplitude of notch and diastolic peak), thereby compromising downstream sensing performance, even in stationary scenarios~\cite{Pi2021PressureOnWaveforms, Charlton_Pilt_Kyriacou_2022, Scardulla2023}. However, \textit{these studies only analyzed the correlation between CP and the AC/DC amplitude, while overlooking the broader morphology of the PPG waveform}. 

We delved into the impact of CP by conducting experiments in real-world conditions. Specifically, we developed a smartwatch-based PPG prototype (\Cref{itw}) to measure the PPG signals from the wrist while a wearer was engaged in different sedentary activities with varying wrist postures. As shown in~\Cref{fig:5Poses}, the PPG waveforms exhibited diverse morphologies under different wrist postures due to the varying CPs. Concretely, we drew several observations: 1) the amplitude difference between the systolic and diastolic peaks fluctuates significantly with changes in CP; 2) the diastolic peak and dicrotic notch are absent in extremely low or high CPs; 3) the ideal waveform is only presented within a proper CP range. The variability in morphology leads to inaccuracies in extracting certain features (e.g., the amplitude and position of the systolic peak or waveform feet), and in some cases, certain features (e.g., diastolic peak and dicrotic notch) may be completely absent. This results in diminished PPG sensing performance or even failure of certain downstream applications. Additionally, we analyzed signal waveforms in four \textit{public} PPG datasets (as introduced in \Cref{sec:eval_downstream_setup}) collected during sedentary conditions and found only 19.6\% of the PPG cycles present the ideal morphology. This reinforces our motivation to enhance suboptimal waveforms since acquiring ideal PPG waveforms is difficult even in experimental, motionless scenarios.

%% file: sections/03-SystemDesign.tex
\section{System Design} \label{systemdesign}

\begin{figure}[t]
    \centering
    \begin{minipage}[t]{0.54\linewidth}
        \centering
        \includegraphics[width=\linewidth]{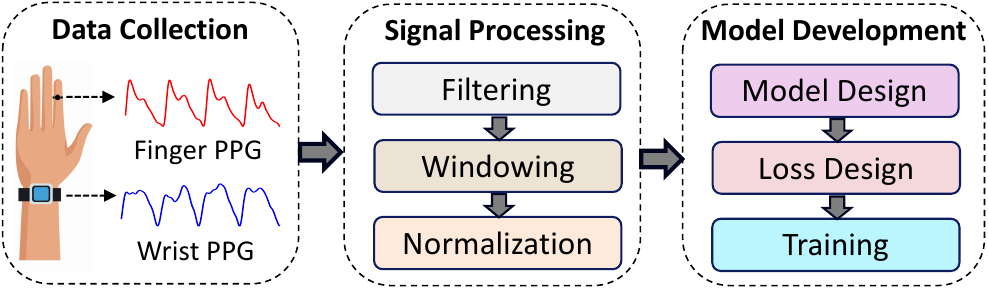}
        \caption{Overview of \SysName Framework.}
        \label{fig:overview}
    \end{minipage}
    \hfill
    \begin{minipage}[t]{0.44\linewidth}
        \centering
        \includegraphics[width=\linewidth]{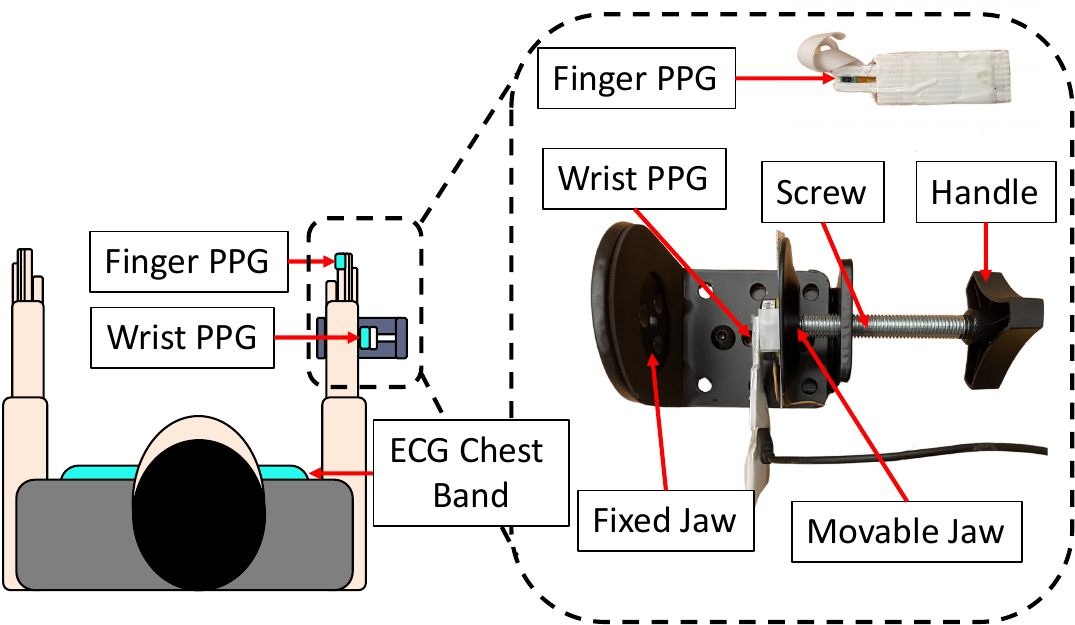}
        \caption{The developed prototype and experiment setup.}
        \label{fig:setupDiagram}
    \end{minipage}
    \vspace{-0.3cm}

\end{figure}

We propose to restore ideal PPG waveforms from distorted ones using a deep learning-based framework. Our approach, as depicted in~\Cref{fig:overview}, consists of three stages. First, we developed a wearable prototype equipped with two PPG sensors to capture signals from both the wrist and the finger for data collection (~\Cref{Prototypinganddataset}). Second, to prepare a high-quality dataset for model training, we devised a comprehensive signal processing pipeline to preprocess the collected PPG data (~\Cref{sec:signal_processing}). Third, we introduced a novel deep neural network along with a set of training strategies to facilitate the training process (~\Cref{sec:DNN}).

\subsection{Prototyping and Data Collection} \label{Prototypinganddataset}

\subsubsection{Motivation}
Restoring the ideal PPG waveform using a deep learning-based approach requires a paired dataset, where the distorted waveform serves as the input and the ideal waveform as the label, for model training. However, most of the existing publicly available PPG datasets contain only a single-stream PPG signal collected from one location. There are a few datasets that collect both wrist and finger PPG simultaneously~\cite{tsai2021coherence, mol2020pulse, beh2021maus}, however, they are designed to address their specific downstream tasks (e.g., feature analysis~\cite{tsai2021coherence}, vasoconstriction~\cite{mol2020pulse}, and mental workload~\cite{beh2021maus}) without specifically concentrating on CP-induced distortions. Our approach, instead, \textit{requires the distorted wrist PPG waveforms to cover a wide range of morphologies that may occur under different CPs}, with the corresponding ideal morphology recorded synchronously. Therefore, we needed to prepare a new, appropriate dataset.

\subsubsection{Prototype Design and Implementation}

Collecting both distorted and ideal PPG simultaneously from the same measurement location is inherently impossible. Therefore, we measure the paired signals from two closely co-located positions: the wrist for distorted waveforms and the index fingertip for ideal waveforms. We chose the wrist because it simulates typical smartwatch usage, where contact pressure variations are common. The rationale for collecting ideal waveforms from the fingertip is threefold: 1) the fingertip is highly perfused and considered an excellent position for PPG acquisition in clinical settings, as in pulse oximeters; 2) the PPG sensor maintains relatively consistent and proper contact pressure with the skin when clipped to the fingertip, increasing the likelihood of capturing the ideal morphology; 3) the proximity of the fingers to the wrist (of the same hand) ensures that both PPG sensors measure blood pulses as similarly. 

Therefore, we designed a wearable prototype comprising two PPG sensors (MAXM86161~\cite{MAXM86161}), referred to as wrist PPG and finger PPG, as shown in~\Cref{fig:setupDiagram}. Both PPG sensors are sampled by MAXSensor BLE~\cite{MAXM86161EVSYS} boards connected to a laptop and the measurements are saved locally as CSV files. To generate various PPG waveform morphologies for the wrist PPG signal, we applied an external stimulus to alter the contact pressure between the sensor and the skin. We achieved this using an adjustable clamp, selected for three key reasons: 1) the experiment instructor can conveniently make fine-grained contact pressure adjustments by turning the clamp's handle, which shifts the screw, movable jaw, and the wrist PPG sensor, thereby producing different PPG morphologies; 2) the fixed jaw of the clamp securely holds the wrist, minimizing natural motion during a measurement session lasting a few minutes; 3) compared to a wristwatch setup (where the strap wraps around the wrist), the clamp applies contact pressure only to the measurement site, reducing the risk of affecting the finger PPG signal where the blood flows from the wrist.

\subsubsection{Data Collection}
Next, we leveraged the developed prototype to collect the required data from human subjects. Specifically, we recruited 22 young and healthy adult volunteers (15 Male, 7 Female, mean age 24.01 $\pm$ 2.33)\footnote{We opted for only young and healthy adults as the elderly or people with cardiovascular diseases may not present the ideal PPG morphology~\cite{fine2021sources, Qawqzehaging}.}. The data collection procedure was reviewed and approved by the University's Institutional Review Board, ensuring that participants provided informed consent and that the experiments were conducted in a comfortable and safe manner.

The procedure consists of three steps: 1) Participants were briefed before the experiment and signed an informed consent form. 2) Participants were instructed to sit still with both arms outstretched on a table. The developed prototype and other devices were then fitted onto the participants' right hands, as illustrated in~\Cref{fig:setupDiagram}. Specifically, the finger PPG sensor was secured to the index finger with a strap to maintain proper contact pressure, similar to a pulse oximeter setup. The participant's wrist was positioned between the fixed and movable jaws of the prototype, with the wrist PPG sensor attached. Additionally, participants wore an ECG chest band (Polar H10~\cite{PolarH10}) to record the ECG signal simultaneously, providing the ground truth for blood pulses. 3) Then, each participant underwent six four-minute data collection sessions, with a two-minute break between sessions. During these breaks, participants were encouraged to move their arms freely to alleviate any fatigue. Additionally, the experiment instructor adjusted the contact pressure of the wrist PPG sensor by turning the screw to produce diverse PPG morphologies. Since the optimal contact pressure varies among individuals due to their unique physiological characteristics, we manually adjusted the contact pressure for each subject in each session to capture a broad spectrum of morphologies across participants. 

In total, we collected $4 \times 6 \times 22 = 528$ minutes of data. It includes three simultaneously recorded signals for dealing with contact pressure variations: 1) wrist PPG signal subject to a full range of contact pressures, 2) ideal PPG signal, and 3) ECG signals. Here, the two PPG signals were sampled at 128~Hz, and the ECG signal at 130~Hz.

\subsection{Signal Processing}
\label{sec:signal_processing}

PPG signals are generally noisy, so to prepare them for model training, we developed a customized suite of signal processing techniques as follows. \Cref{fig:preprocessing} illustrates the signals at different stages of the processing pipeline:
\begin{figure}[h] 
    \vspace{-0.3cm}
    \centering\includegraphics[width=\linewidth]{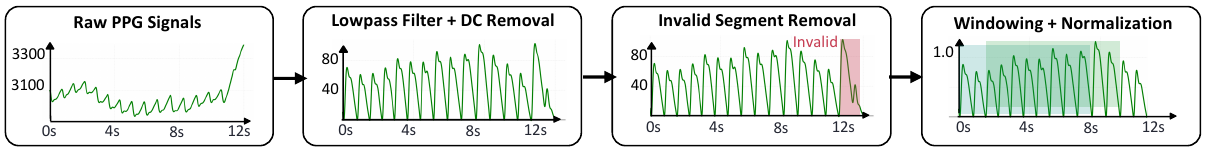}
    \vspace{-0.24in}\caption{Illustration of PPG signals after applying different processing techniques.}\vspace{-0.1in}
    \label{fig:preprocessing}
\end{figure}

\textit{Lowpass filtering: }PPG signals usually contain high frequency noise caused by various factors, such as ambient light variations and electronic noise~\cite{Fine2021NoiseSources}. Therefore, similar to existing literature~\cite{Park2022SurveyPaper}, we applied a 5th-order Butterworth lowpass filter with a cutoff frequency of 10~Hz to the two PPG streams to remove high frequency signal components.

\textit{DC removal: } This study focuses on restoring the missing PPG features related to pulsatile blood flow, which strongly pertains to the AC component as discussed in \Cref{sec: pre_downstream}. Thus, we extracted the AC component by removing the DC component. Specifically, we began by using peak detection to identify signal troughs, which were then used to derive the DC component through cubic spline interpolation. The AC component was then isolated by subtracting this DC baseline from the filtered PPG signals.

\textit{Invalid segment removal: } Although participants were instructed to remain still during each four-minute session, occasional wrist and finger movements were unavoidable. Consequently, some finger PPG waveforms deviate from the ideal morphology, and wrist PPG signals may be corrupted by motion artifacts. Additionally, inherent heartbeat variations, influenced by factors such as breathing, can cause deviations in PPG waveforms~\cite{Hartmann2019BreathingAffectsWaveforms}. Given that the finger PPG serves as the ground truth for model training, it is crucial to ensure all finger PPG signals used for training exhibit the ideal morphology. To achieve this, we calculated the similarity between each detected PPG cycle and the ideal waveform, removing segments with similarity scores below a threshold (set to ensure the presence of dicrotic notch, systolic, and diastolic peak). Corresponding segments in the wrist PPG signal were also eliminated to maintain synchronization. Consequently, finger PPG signals exhibit ideal morphology, while wrist PPG retains diverse patterns.

\textit{Windowing: } Next, to facilitate the model training, we split the signals into windows using the sliding window technique. Following existing studies~\cite{Zhang2015EightSecondWindows,mehrgardt2022pulse,misc_ppg-dalia_495}, the default window length was designated as 8 seconds, with 1.6 seconds sliding increment. We evaluate the impact of window length on the transformation performance in~\Cref{Window_Length}. After windowing, the wrist and finger PPG signals were converted to multiple instances, each comprising a pair of synchronized windows derived from the two signals.

\textit{Normalization: } PPG signal amplitudes can vary greatly among individuals due to factors such as skin tone variations~\cite{fine2021sources}, strengths of heartbeats~\cite{Scardulla2023}, and even within an individual owing to factors like contact pressure~\cite{SIRKIA2024PPGandCP}. Since we aim to restore the morphology of the PPG signal, i.e., relative amplitudes within a window, we applied the max-min normalization to scale each window between 0 and 1.

\textit{Data augmentation: }Due to varying participant characteristics, behaviors (e.g., different frequencies and degrees of hand movements), and skin factors, the percentage of valid segments across individuals can be uncontrollably discrepant. We balanced the dataset using data augmentation techniques, including random time warping, jittering, baseline drifting, and amplitude scaling~\cite{10.1007/s00521-023-08459-3}. Ultimately, we obtained over 4,600 eight-second [wrist PPG, finger PPG] window pairs for subsequent model training and evaluation.

\subsection{Deep Neural Network Design} \label{sec:DNN}

\begin{figure}[t] 
    \centering\includegraphics[width=0.7\linewidth]{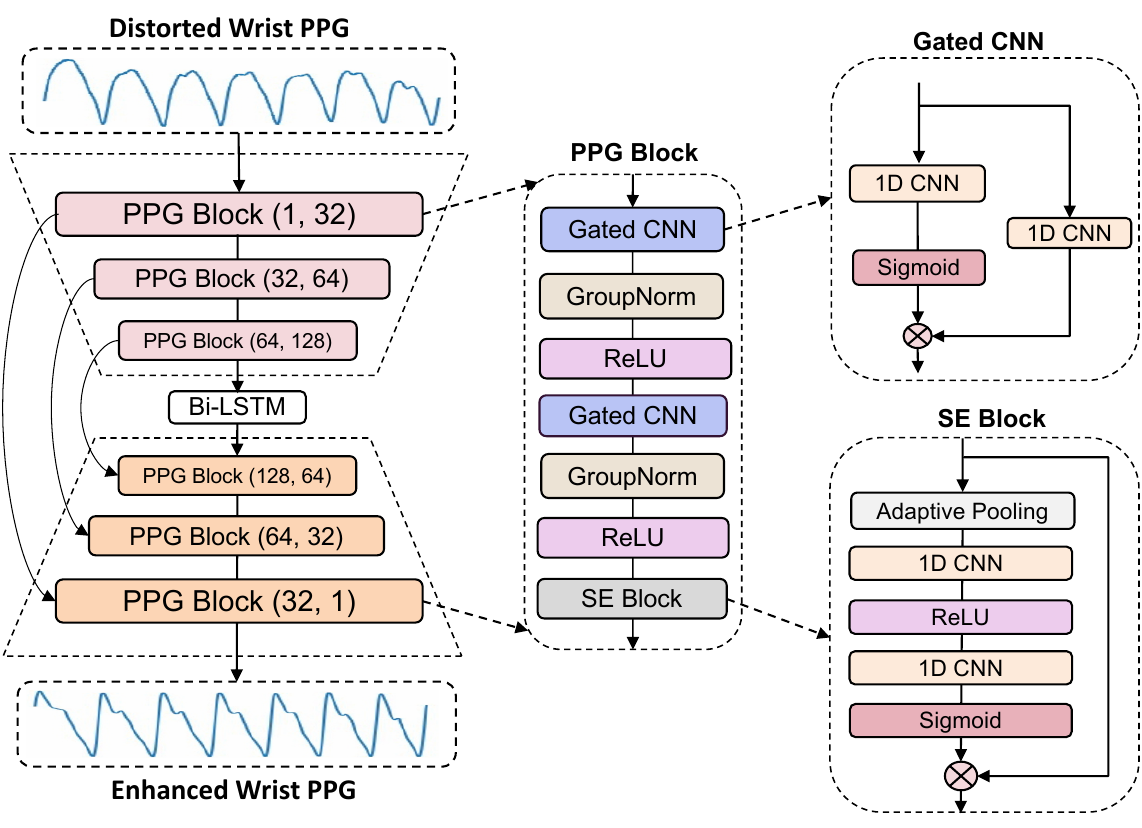}
    \vspace{-0.15in}\caption{Illustration of the generator architecture of the \SysName model.}
    \label{fig:generator}
\end{figure}

To effectively transform the distorted PPG morphology into the ideal morphology, we propose a lightweight generative adversarial-based architecture as presented in~\Cref{fig:generator}. At a high level, it is composed of a generator ($G$) sub-network and a discriminator ($D$) sub-network, and was trained in an adversarial manner. The generator transforms the distorted wrist PPG signals into high-quality signals, while the discriminator functions as an auxiliary component which, together with a custom loss function (\Cref{sec:custom_loss}), further enhances the generator's capacity to generate better-fidelity samples. 

\subsubsection{Generator} For the generator model, we designed an enhanced autoencoder model with a fully convolutional encode–decode architecture and a Bi-directional Long Short-Term Memory layer (Bi-LSTM) introduced in the latent space. The encoder and decoder include three PPG blocks with skipping connections between corresponding blocks at different feature resolutions. Each PPG block starts with two sequences of Gated Convolutional Neural Network (Gated CNN) followed by a GroupNorm, a regular ReLU activation, and ends up with a Squeeze and Excitation (SE) Block.  In the encoder, the number of channels initially starts at 1 and progressively increases to 32, 64, and finally 128, as the input signal passes through the convolutional layers within the PPG Blocks to get high-level features from the input. Conversely, the number of channels in the decoder gradually decreases, reverting to 1 as the desired enhanced signal. Meanwhile, the Bi-LSTM layer in the latent space enhances the model's ability to capture temporal and long-range dependencies. It’s worth noting that this layer will double the number of output channels. Therefore, we incorporate a Dense layer to reduce it back to 128. In the following sections, we further detail important components of the generator model.

\noindent\textbf{Gated CNN Block}. The Gated CNN block is inspired by the GRU layer's gate design \cite{cho2014learning}, which can be used to retain necessary features while reducing unimportant input features. As shown in~\Cref{fig:generator}, our Gated CNNs are designed as a 1D CNN layer with a kernel size of 3, a stride of 1, and a padding of 1, adapting an additional gating mechanism. On one hand, the CNN layer aids the model in learning generalized features from dynamically varying PPG signals, allowing the encoder and decoder to capture useful temporal dependencies. On the other hand, the gating mechanism adjusts the flow of information through another 1D CNN layer using the Sigmoid function, which emphasizes/suppresses aspects of the signal based on their relevance to the learning context.

\noindent\textbf{Squeeze and Excitation (SE) Block}. Instead of assigning equal importance to all embedding channels as in traditional CNN layers, the SE block dynamically adjusts the weights of each channel based on its importance in capturing relevant features during the learning process. In our proposed model, it comprises two primary stages: squeezing and excitation, as shown in~\Cref{fig:generator}. In the squeezing stage, the network first employs a global average pooling layer where the mean value of every channel is calculated, resulting in a global context vector. This context vector is processed through two successive convolutional layers, with ReLU activation in between. The first convolutional layer reduces the channel by twice, while the second restores it to the original size. Finally, a Sigmoid function is employed to output attention weights in the range between 0 and 1. In the excitation stage, a residual connection from the input goes to a dot product with those weights from the squeezing stage to modulate the importance of each channel in the original input. 

\noindent\textbf{Bi-LSTM Layer}.
The Bi-LSTM layer aims to support, and comprehend the temporal dynamics and spatial hierarchies present in the PPG signals. The input features from the encoder are processed both forward and backward in the Bi-LSTM layer to identify long-term contextual characteristics and dependencies of the signals. Specifically, the PPG cycles exhibit temporal correlations and shared patterns in a given window. As a result, using Bi-LSTM here enables richer spatial-temporal feature representation in the latent space, improving the understanding of the signal during the transformation.

\subsubsection{Discriminator} 
We further apply an adversarial training approach with the help of a discriminator model to enhance our model's robustness. It is designed with the same architecture as the encoder from the proposed generator model architecture, except for one additional Dense layer in the end to output a single value. During adversarial training, the generator is trained to minimize the Hinge loss~\cite{gentile1998linear}, which is widely used in the context of adversarial training. It encourages the generator to produce PPG signals that are indistinguishable from reference signals by the discriminator. Simultaneously, the discriminator is trained to minimize the Hinge loss for real samples and maximize it for generated samples, leading to a more robust discriminator that can better distinguish between real and generated data. The Hinge loss function used by the discriminator is:

\begin{equation}
L_D = \frac{1}{N} \sum_{i=0}^{N} [\max(0, 1 - D(x_i)) + \max(0, 1 + D(z_i))],
\end{equation}
\noindent where \( N \) is the number of windows, \( x_i \) is the reference signal, and \( z_i = G(y_i) \) is the enhanced signal, with \( y_i \) as the original signal. \( D(x_i) \) and \( D(z_i) \) are the discriminator's predictions for the reference and enhanced signals, respectively.

\subsubsection{Custom Loss Function} 
\label{sec:custom_loss}
In this section, we introduce a custom loss function specifically designed for our task. Existing approaches for reconstructing corrupted signals have primarily relied on mean squared error (MSE) loss. However, due to the high sensitivity of PPG signals to environmental factors and amplitude variations, models trained with MSE alone often become unstable and struggle with generalization~\cite{wang2009mean}. Moreover, a key limitation of MSE is that it treats all errors uniformly, regardless of their position within the signal. This is problematic because certain components of the PPG signal, such as the systolic peak, dicrotic notch, and diastolic peak, are more critical than others. To address this, we propose a custom loss function that incorporates PPG-specific domain knowledge, as shown in~\Cref{eq:cusom-loss}. 

\begin{equation}
L_C =
\begin{cases}

MSE(z_i, x_i) \cdot \alpha + MSE(z_i, x_i) & \text{if } \alpha \neq 0, \\ 

L_{P2P}(z_i, x_i) \cdot \beta + MSE(z_i, x_i) & \text{if } \alpha = 0 

\end{cases}
\label{eq:cusom-loss}
\end{equation}

Specifically, based on the premise that an ideal PPG cycle is characterized by two peaks (systolic and diastolic peak) and one notch (dicrotic notch) between them, we introduce a variable, $\alpha$, which represents the difference in the number of peaks between the estimated and reference signals. Then, our custom loss function operates in two stages: 1) In the early stages of training, when the model often predicts PPG signals with an incorrect number of peaks ($\alpha \neq 0$), we use an additional term to penalize this by applying a factor proportional to $\alpha$ to the MSE loss. \textit{This encourages the model to produce PPG cycles with the correct number of peaks (i.e., 2)}. 2) As training progresses and the PPG cycles approach the correct number of peaks (i.e., $\alpha=0$), we introduce an additional point-to-point loss ($L_{P2P}$) term scaled by a factor $\beta$ ($0.01$ empirically) over the MSE loss. Concretely, $L_{P2P}$ is calculated as the amplitude difference of the three points (two peaks and one notch) between the estimated and reference signals, which \textit{encourages the model to refine the relative amplitude of the estimated signal}. Guided by this custom loss function, the model is better equipped to predict accurate and ideal PPG waveforms. Eventually, we define the generator loss as the combination of custom loss and generator-side Hinge loss, as follows:

\begin{equation}
\label{eq:loss}
L_G = \frac{1}{N} \sum_{i=0}^{N} L_C - \theta \cdot \frac{1}{N} \sum_{i=0}^{N} D(z_i),
\end{equation}
\noindent where $\theta$ is a hyper-parameter that is set to be $0.01$ empirically and $N$ denotes the total number of windows. 

\subsubsection{Model Training}
We partitioned the 22 participants into three separate sets: training (13 participants), validation (4 participants), and testing (5 participants). Each instance consists of eight-second wrist PPG and finger PPG pairs. \SysName was implemented in PyTorch and trained on a single NVIDIA GeForce RTX 3090 GPU. The AdamW optimizer with $\beta_1 = 0.9$, $\beta_2 = 0.999$ was utilized for training. The initial learning rate was set to $0.001$ and a learning rate scheduler with an $l2$ regularization term with a scale of $0.005$ was applied. We limited the maximum training epoch to 300 and implemented the early stopping mechanism on the validation set to prevent model overfitting.

%% file: sections/06-PerformanceEvaluation.tex
\section{Evaluation} \label{evaluation}

To evaluate \SysName, we conducted comprehensive experiments as follows: 1) First, using the self-prepared dataset, we assessed the performance of the deep learning model from a morphological perspective by examining various waveform features. 2) Subsequently, we examined the effectiveness of the pre-trained model on four typical PPG downstream tasks by comparing the sensing performance obtained with PPG signals before and after the model transformation, on both the self-prepared dataset and other public PPG datasets. 3) Finally, we upgraded our prototype in the form of a wristwatch and conducted a longitudinal in-the-wild test to validate the effectiveness of \SysName in real-life scenarios.

\subsection{Waveform Transforming Performance}

\subsubsection{Metrics}

We used several metrics to evaluate the transformation performance from different perspectives. First, we considered three \textbf{overall metrics}: Mean Absolute Error (MAE), Dynamic Time Warping (DTW), and Pearson Correlation Coefficient (PCC). MAE measures the average error magnitude between the enhanced wrist PPG and finger PPG signals, reflecting overall accuracy. DTW accounts for temporal variations and optimally aligns the signals, offering a detailed comparison of their morphological similarity. PCC indicates the strength of correlation between the enhanced wrist PPG and reference finger PPG, with higher values showing greater waveform similarity.

\begin{figure}
    \centering
    \begin{minipage}{0.33\linewidth}
        \centering
        \includegraphics[width=0.79\linewidth]{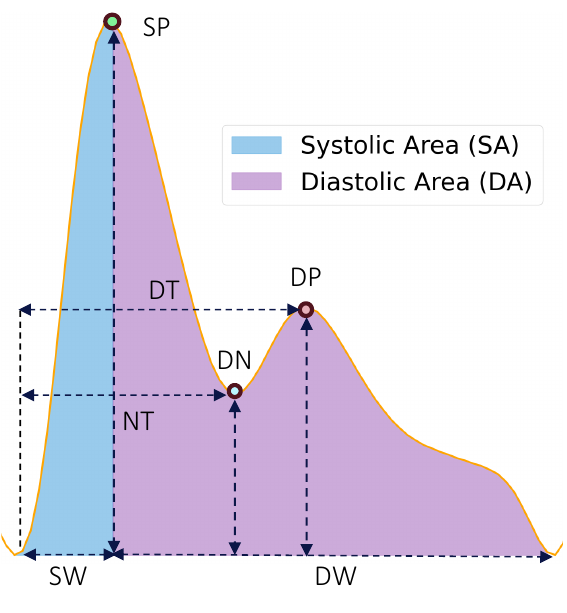}
        \vspace{-0.1in}
        \caption{PPG point-based and area-based features used for \SysName's evaluation.}
        \label{fig:features}
    \end{minipage}
    \hfill
    \begin{minipage}{0.66\linewidth}
        \centering
        \includegraphics[width=0.99\linewidth]{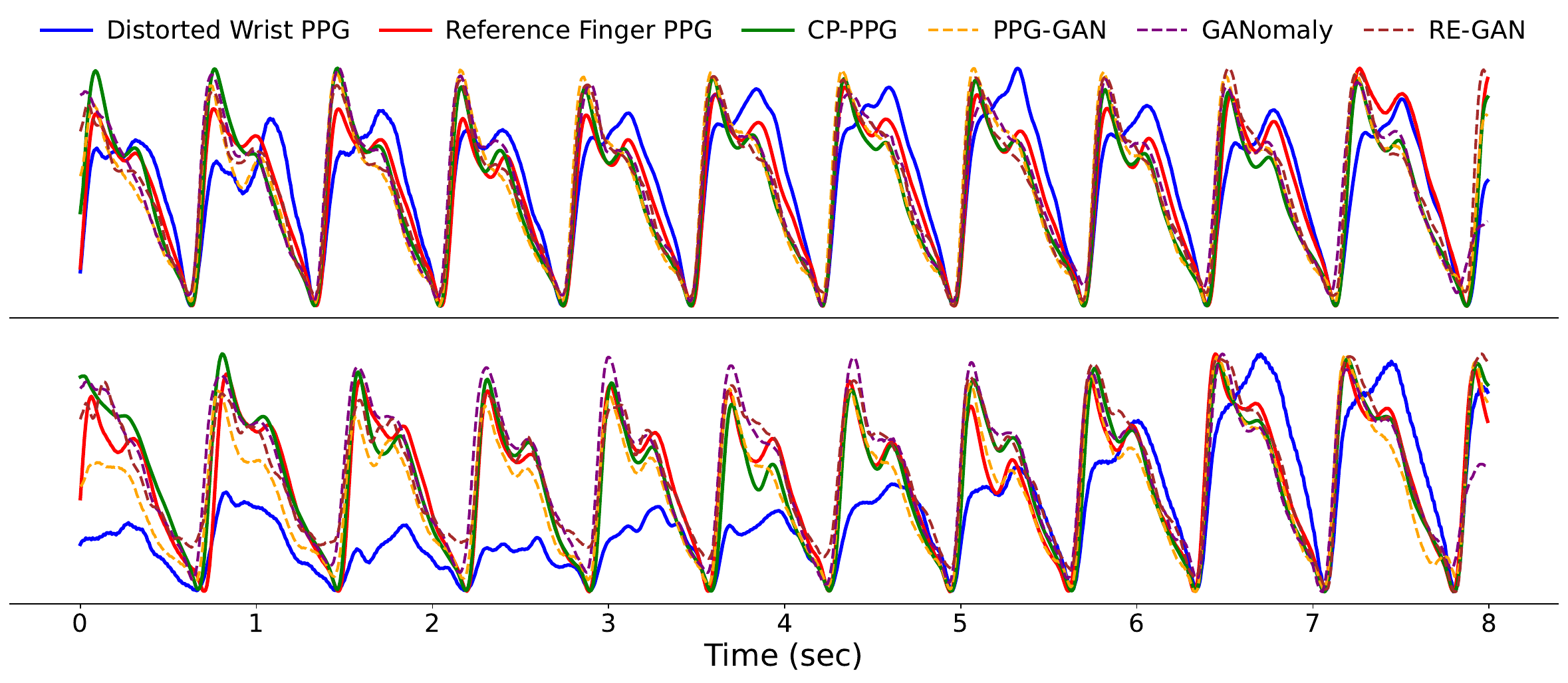}
        \vspace{-0.1in}
        \caption{Two examples illustrate the effectiveness of \SysName in refining the distorted wrist PPG signal.}
        \label{fig:output_waveforms}

    \end{minipage}
\end{figure}

Then, we further investigate detailed PPG waveform features, including point-based, area-based, and signal quality index (SQI)-based metrics. As shown in ~\Cref{fig:features}, an ideal PPG cycle comprises three essential fiducial points: the systolic peak, diastolic peak, and dicrotic notch situated between them with their amplitudes and time scales are considered crucial features for quantifying the morphology of a PPG waveform and hold significant clinical importance. Thus, we considered \textbf{point-based metrics}~\cite{Chowdhury2020bpWithML, gonzalez2023benchmark} including: the amplitude of the systolic peak (SP), time span from onset to the systolic peak (SW), time span from systolic peak to the next onset (DW); the amplitude of the dicrotic notch (DN), time span from onset to the dicrotic notch (NT); the amplitude of the diastolic peak (DP), time span from onset to the diastolic peak (DT). We utilized the mean absolute percentage error (MAPE) between the enhanced and reference PPG signals to demonstrate the transformation performance. 

In addition to point-based metrics, areas under the curve of the signal also provide valuable insights into vascular dynamics~\cite{gonzalez2023benchmark}. Therefore, we consider two \textbf{area-based features}: systolic area (SA) and diastolic area (DA), which are defined based on the position of the systolic peak. We also used MAPE for area-based metrics. 

Finally, we investigate \SysName's effectiveness using \textbf{signal quality indexes} (SQIs) including skewness and kurtosis, which are commonly used due to their ability to provide valuable insights into the shape and distribution of PPG waveform~\cite{elgendi2016optimal}. Specifically, in typical PPG signals, a slightly positive skewness is common, as the systolic peak is usually more pronounced, suggesting normal blood flow dynamics. Similarly, kurtosis measures the peakness or tailedness of a distribution, reflecting the shape of the waveforms and providing insights into their asymmetry. PPG signals often exhibit moderate kurtosis, reflecting the sharp systolic peak characteristic of normal waveforms. Here, we report skewness and kurtosis errors as Mean Absolute Error (MAE). 

\subsubsection{Baselines}
In recent years, researchers have explored adversarial-based methods for PPG signal reconstruction using deep neural networks. For example, PPG-GAN~\cite{Zheng2022_6} employs a traditional Pix2Pix-inspired network~\cite{isola2017image} with an U-Net-based generator to reconstruct motion-distorted PPG signals. Building upon this, RE-GAN~\cite{long2023reconstruction} integrates a CycleGAN-based generator~\cite{zhu2017unpaired} with a Pix2Pix-based discriminator, further enhanced by a recursive loss to ensure consistency in the generated signals. Another approach~\cite{wang2022ppg} adopts the GANomaly-based model, incorporating an additional latent space loss to align the embeddings of the generated and reference signals, thereby improving reconstruction performance. These methods primarily address motion artifacts during the reconstruction phase and aim to recover coarse-grained PPG waveforms (e.g., restoring the systolic peak). In contrast, \SysName focuses on reconstructing the fine-grained morphology of PPG signals, preserving all key features in the ideal morphology. We compare our method against the three aforementioned approaches, using the same dataset for both training and inference.

\subsubsection{Main Results}

\begin{table}[h]
\centering
\scriptsize
\setlength\tabcolsep{5pt}
\vspace{-0.1in}

\caption{Waveform transformation performance with overall metrics (MAE, DTW, PCC), point-based metrics (MAPE \%), area-based metrics (MAPE \%), and SQI-based metrics (MAE).}
\vspace{-0.1in}
\label{tab:transform_performance}
\begin{tabular}{lccccccccccccccccccc}
\toprule
& \multicolumn{3}{c}{\textbf{Overall Metrics}} & \multicolumn{7}{c}{\textbf{Point-based Metrics}}  & \multicolumn{2}{c}{\textbf{Area-based Metrics}} & \multicolumn{2}{c}{\textbf{SQI-based Metrics}} \\
\cmidrule(lr){2-4} \cmidrule(lr){5-11} \cmidrule(lr){12-13} \cmidrule(lr){14-15}
& MAE & DTW & PCC & SP & SW & DW & DN & NT & DP & DT & SA & DA & Skewness & Kurtosis \\
\midrule
Distorted & 0.15 & 0.14 & 0.77 & 19.92 & 37.98 & 8.90 & 25.83&8.19 &32.85&10.32&49.02&19.23&0.42&0.14\\

PPG-GAN~\cite{Zheng2022_6} & 0.10 & 0.09 & 0.89 & 10.21&20.96&5.01&13.35&10.01&16.45&12.08&25.08&16.23&0.27&0.13 \\

GANomaly~\cite{wang2022ppg} & 0.11 & 0.10 & 0.87 & 17.40 & 18.23 & 4.63 & 14.48 &13.21 &18.29&11.67&21.75&20.08&0.22&0.13  \\

RE-GAN~\cite{long2023reconstruction} & 0.10 & 0.10 & 0.88 & 11.92 & 17.47 & 4.37 &15.21&11.10&20.52&9.87&22.32&13.97&0.24&0.12 \\

\textbf{\SysName} & \textbf{0.09} & \textbf{0.06} & \textbf{0.91} & \textbf{6.82} & \textbf{16.68} & \textbf{4.08} & \textbf{10.03} & \textbf{5.51} & \textbf{11.32} & \textbf{6.74} & \textbf{13.59} & \textbf{9.21} & \textbf{0.19} & \textbf{0.12} \\

\bottomrule
\vspace{-0.2in}
\end{tabular}
\end{table}

\textit{\\Waveform Illustration}:  \Cref{fig:output_waveforms} illustrates two examples of the distorted wrist PPG, reference finger PPG, and enhanced wrist PPG generated by three baselines and \SysName. Overall, the distorted wrist PPG signals are transformed to closely resemble the ideal morphology of the reference finger PPG signals, when using \SysName. 
This superior performance can be attributed to \SysName's unique design, which incorporates specialized modeling components, such as the custom PPG-aware loss function. Notably, we found that 96.12\% of PPG cycles lacking the diastolic peak were successfully recovered by \SysName. In contrast, while the three baselines also restore the coarse-grained morphology, they perform poorly in recovering the diastolic peaks and dicrotic notches, which are essential for certain downstream tasks.

\textit{Quantitative Results: }
\Cref{tab:transform_performance} presents a quantitative comparison of \SysName with other baselines on the aforementioned metrics. The results clearly show that \SysName achieves significantly lower errors across all metrics compared to the distorted case, demonstrating its ability to accurately recover and refine the location and amplitude of the three essential fiducial points. While the baselines also show some improvement, their performance remains consistently inferior to \SysName, which excels in handling the CP variations that the baselines are not specifically designed to address.

These quantitative analyses, complemented by visual comparisons (\Cref{fig:output_waveforms}), highlight the effectiveness of \SysName in rectifying distorted wrist PPG signals, aligning them more closely with the ideal characteristics of finger PPG signals, thereby potentially enhancing the sensing performance of downstream physiological tasks. 

\subsection{Ablation Studies}
In this section, we evaluate the impact of different model configurations on transformation performance to gain insight into how \SysName achieves its results.

\subsubsection{Impact of Model Components} To assess the contribution of different model components, including the PPG blocks, Bi-LSTM layers, custom loss function, and adversarial training scheme, we systematically removed one component at a time from the default proposed model. Specifically, we started by eliminating the adversarial training scheme, which removes the discriminator and trains the model with the generator only. Subsequently, the proposed custom loss function was removed and replaced by normal MSE loss. Next, we took the Bi-LSTM layer away to assess its effectiveness in capturing long-term dependencies in the model's latent space. Lastly, by replacing the designed PPG blocks with standard CNN blocks of the same depths, we examined a normal 1D CNN encode-decode model. 

\begin{table}[h]
\centering
\scriptsize
\caption{Impact of model components: \textcircled{1} PPG blocks, \textcircled{2} Bi-LSTM, \textcircled{3} Custom loss, \textcircled{4} Adversarial training.}
\vspace{-0.1in}
\label{tab:model_components}
\setlength\tabcolsep{5pt}
\begin{tabular}{@{}cccccccccccccccccc@{}}
\toprule

\multicolumn{4}{c}{\textbf{Components}} & \multicolumn{3}{c}{\textbf{Overall Metrics}} & \multicolumn{7}{c}{\textbf{Point-based Metrics}} & \multicolumn{2}{c}{\textbf{Area-based Metrics}} & \multicolumn{2}{c}{\textbf{SQI-based Metrics}} \\

\cmidrule(lr){1-4} \cmidrule(lr){5-7} \cmidrule(lr){8-14} \cmidrule(lr){15-16} \cmidrule(lr){17-18}

\textcircled{1} & \textcircled{2} & \textcircled{3} & \textcircled{4}  & MAE & DTW & PCC & SP & SW & DW & DN & NT & DP & DT & SA & DA & Skewness & Kurtosis\\ 
\midrule

\textbf{$\checkmark$}& \textbf{$\checkmark$}& \textbf{$\checkmark$}& \textbf{$\checkmark$} & \textbf{0.09} & \textbf{0.06} & \textbf{0.91} & \textbf{6.72} & \textbf{16.68} & \textbf{4.08} & \textbf{10.03} & \textbf{5.51} & \textbf{11.32} & \textbf{6.74} & \textbf{13.59} & \textbf{9.21} & \textbf{0.19} & \textbf{0.12} \\ 

\midrule

$\checkmark$    &$\checkmark$&$\checkmark$& & 0.09 & 0.07 & 0.90 & 6.82 & 16.11 & 4.37 & 11.35 & 5.37 & 12.09 & 6.81 &15.61 &12.84 &0.25 & 0.16  \\
$\checkmark$    &$\checkmark$& & & 0.10 & 0.08 & 0.88 & 6.88 & 18.04 & 4.48 & 11.31 & 6.15 & 12.56 & 8.20 & 17.51 & 16.37 &0.33&0.19 \\
$\checkmark$    & & & &0.11 &0.08 &0.87 & 6.92 & 19.35 & 4.86 & 12.81 & 6.39 & 14.58 & 8.20 &21.73&17.01 & 0.35 & 0.19 \\
& & & & 0.13& 0.09& 0.85 & 8.36 & 18.54 & 4.63 & 20.70 & 6.55 & 22.48 & 8.96 &19.68&22.33&0.39& 0.24 \\

\bottomrule
\vspace{-0.2in}
\end{tabular}
\end{table}

~\Cref{tab:model_components} displays the settings and their corresponding performance in overall metrics and our predefined feature-based metrics. We can observe that the transformation performance gradually decreases with the removal of each additional component, suggesting that every component positively contributes to the model. Notably, the MAPEs of the diastolic-related metrics suffer from a significant increase after removing the custom loss function and adversarial training setting, which in turn demonstrates their crucial roles in the model. This is attributed to the special design of the custom loss, which encapsulates the intricate domain knowledge inherent in PPG waveforms, compelling the model to prioritize the accurate restoration of critical features like the diastolic peak. Meanwhile, in adversarial training, the discriminator robustly learns to distinguish between the enhanced and reference signals, in a way that guides the generator toward producing outputs that more closely resemble authentic signals.

\subsubsection{Impact of Model Depth (MD)} 

The PPG block, designed to capture the intricate characteristics of the PPG waveform, has proven to be highly effective in improving the transformation performance of \SysName, as evidenced in~\Cref{tab:model_components}. In the default model, we stacked three PPG blocks in the encoder and decoder, respectively. Since the model depth ( i.e., the number of PPG blocks) directly influences the model's capacity, we evaluate its impact on the transformation performance. Specifically, we trained two additional models with $MD$=1 and $MD$=5 respectively, and compared their performance with the default model ($MD$=3).

\begin{table}[h]
\centering
\scriptsize
\vspace{-0.1cm}
\caption{Impact of model depth (MD).}
\label{tab:model_depth}
\vspace{-0.2cm}
\setlength\tabcolsep{4.5pt}
\begin{tabular}{@{}cccccccccccccccccc@{}}
\toprule

& \multirow{2}{*}{\textbf{MD}} & \multirow{2}{*}{\textbf{\#Params}} & & \multicolumn{3}{c}{\textbf{Overall Metrics}} & \multicolumn{7}{c}{\textbf{Point-based Metrics}}  & \multicolumn{2}{c}{\textbf{Area-based Metrics}}  & \multicolumn{2}{c}{\textbf{SQI-based Metrics}}  \\ 

\cmidrule(lr){5-7} \cmidrule(lr){8-14} \cmidrule(lr){15-16} \cmidrule(lr){17-18}

& & & & MAE & DTW & PCC & SP & SW & DW & DN & NT & DP & DT & SA & DA & Skewness & Kurtosis \\ 
\midrule

& 1 & 80K & & 0.11& 0.08 & 0.89& 8.15 & 17.11 & 4.93 & 11.90 & 6.61 & 12.03 & 7.73 &14.31&14.04&0.27&0.13 \\ 
&3 & 1.1M & & \textbf{0.09} & \textbf{0.06} & \textbf{0.91} & \textbf{6.82} & \textbf{16.68} & \textbf{4.08} & \textbf{10.03} & \textbf{5.51} & \textbf{11.32} & \textbf{6.74} & \textbf{13.59} & \textbf{9.21} & \textbf{0.19} & \textbf{0.12} \\

&5 & 17.5M &&0.08 &0.06 & 0.91 & 7.48 & 15.62 & 3.79 & 9.77 & 5.39 & 11.48 & 7.36 &12.16&9.48&0.20&0.12 \\

\bottomrule
\end{tabular}
\end{table}

As shown in~\Cref{tab:model_depth}, the transformation performance tends to increase with the model depth, attributed to the increased model capacity. However, we observed a marginal improvement in performance or even degradation on specific features (e.g., SP, DT) when $MD$ increases from 3 to 5. This could be attributed to the model with $MD=3$ already capturing the correlation between waveform transformation, and deeper models might lead to overfitting. The model with $MD=5$ (17.5M parameters) is also approximately 16 times larger than that with $MD=3$ (1.1M parameters), significantly increasing the system overhead (e.g., latency and power consumption) of training and inference. Therefore, we selected $MD=3$ as the default setting as it strikes an optimal balance between transformation performance and system overhead.

\subsubsection{Impact of Input Window Length (WL)} \label{Window_Length}

In the default configuration, we segmented the PPG signals into eight-second windows for model training and testing. As the PPG signal presents strong temporal correlations, the length of the input window might affect the transformation performance. Thus, for this ablation study, we prepared two additional versions by segmenting the PPG signals into windows with lengths of three seconds and five seconds, respectively, which are then utilized to train the corresponding models. Notably, this experiment does not extend to longer window lengths as it results in fewer windows, which can weaken our model training.

 \begin{table}[h]
\scriptsize
\vspace{-0.2cm}
\caption{Impact of input window length (WL).}
\vspace{-0.1in}
\label{tab:window_length}

\setlength\tabcolsep{5pt}
\begin{tabular}{@{}cccccccccccccccccc@{}}
\toprule

& \multicolumn{2}{c}{\textbf{WL}} & & \multicolumn{3}{c}{\textbf{Overall Metrics}} & \multicolumn{7}{c}{\textbf{Point-based Metrics}} &\multicolumn{2}{c}{\textbf{Area-based Metrics}} &\multicolumn{2}{c}{\textbf{SQI-based Metrics}}   \\ \cmidrule(lr){1-4} \cmidrule(lr){5-7} \cmidrule(lr){8-14} \cmidrule(lr){15-16} \cmidrule(lr){17-18}
& Train & Test & & MAE & DTW & PCC & SP & SW & DW & DN & NT & DP & DT & SA & DA & Skewness & Kurtosis\\ 
\midrule
& 3s & 3s & &0.12 & 0.12& 0.85& 8.41  & 17.53 & 4.26 & 13.14 & 6.84 & 17.79 & 8.04 &17.22&11.15&0.22&0.16 \\
& 5s & 5s & & 0.12& 0.09 & 0.87& 7.15  & 16.79 & 4.07 & 11.05 & 6.47 & 14.26 & 7.07 & 16.93&11.49 &0.20&0.14 \\
& \textbf{8s}& \textbf{8s} & & \textbf{0.09} & \textbf{0.06} & \textbf{0.91} & \textbf{6.82} & \textbf{16.68} & \textbf{4.08} & \textbf{10.03} & \textbf{5.51} & \textbf{11.32} & \textbf{6.74} & \textbf{13.59} & \textbf{9.21} & \textbf{0.19} & \textbf{0.12} \\
\midrule

& 8s &  3s & &0.11 &0.12 & 0.87& 7.89 & 18.40 & 4.16 & 13.04  & 7.01 & 16.37  & 8.74 &17.43&11.60&0.23&0.16 \\
& 8s  & 5s & &0.10 &0.08 & 0.89& 7.01 &  17.19  & 4.13 & 11.13 &  6.06 & 13.39 & 7.05 &15.50&9.95&0.20&0.14\\

\bottomrule

\end{tabular}
\end{table}

From~\Cref{tab:window_length}, we can observe that with the increase in window length, the transformation performance consistently improves across most predefined metrics, suggesting that a larger input window length allows the model to capture the temporal dependencies more accurately. Next, given that our generator with autoencoder-based architecture can accommodate inputs of varying lengths, we further tested the eight-second model using data with three-second and five-second window sets (i.e., only perform inference). Remarkably, the eight-second model still outperformed the other two models, regardless of the input window length during testing. This finding underscores the versatility of our model, which can be effectively deployed across diverse practical scenarios to accommodate
different window sizes.

\subsection{Performance on Physiological Tasks}
\subsubsection{Experiment Setup}
\label{sec:eval_downstream_setup}

\begin{figure}[h]
  \centering  \includegraphics[width=0.6\textwidth]{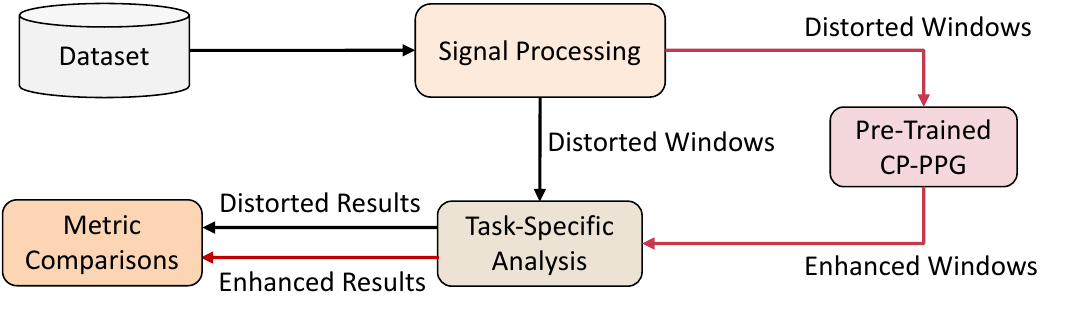} 
  \caption{The evaluation pipeline for PPG downstream tasks. 'Pre-trained \SysName' serves as an external API.}

  \label{fig:downstream_pipeline}
  \vspace{-0.3cm}
\end{figure}

\textit{While the above evaluations demonstrated the effectiveness of \SysName in refining the morphology of PPG waveforms towards ideal ones, it remains uncertain whether the refined waveforms can improve PPG sensing performance and to what extent. }Therefore, we selected four typical PPG downstream tasks - HR estimation, heart rate variability (HRV) estimation, BP estimation, and respiratory rate (RR) estimation - to assess the effectiveness of \SysName using the evaluation pipeline depicted in~\Cref{fig:downstream_pipeline}. 

Specifically, for a given dataset of a certain task, we first applied typical signal processing (e.g., filtering, DC removal, etc.) and segmented the signal into windows with the length required for the task following existing literature (\textit{Distorted Windows}). Then, we fed these windows into the task-specific analysis pipeline (e.g., HR estimation algorithm/model) to estimate the corresponding health parameter, yielding \textit{Distorted Results} by comparing them with the ground truth. Then, we also input the distorted windows into our \textbf{pre-trained} \SysName model for waveform transformation to obtain enhanced windows (\textit{Enhanced Windows}). These windows then underwent the same procedures for task-specific analysis to obtain \textit{Enhanced Results}. Finally, we compared the enhanced and distorted results to understand the effectiveness of \SysName.

\subsubsection{Datasets} We utilized both our self-collected data and public datasets to evaluate \SysName's performance on different downstream tasks. In the self-collected data, we recorded the ECG signal alongside PPG signals, which enables the derivation of the ground truth for HR and HRV. The majority of the PPG datasets in the public domain were acquired during various intensive activities such as walking and running, as they aim to investigate the impact of motion artifacts. On the other hand, since \SysName is designed to address signal distortion resulting from changes in tightness under sedentary conditions, our primary criterion for selecting public datasets was that they should contain PPG data collected in stationary or sedentary scenarios. Following that, we selected four datasets to evaluate different downstream tasks described below. 

\textbf{PTT-PPG}~\cite{mehrgardt2022pulse}: this dataset mainly consists of three data streams, including one ECG and two reflectance PPGs, one from the fingertip and the other from the base of the finger, from 22 healthy subjects while still, walking, and running. We only used the data recorded during stationary periods and enhanced the finger base PPG using pre-trained \SysName to evaluate its performance on HR and HRV estimation.

\textbf{PPG-DaLiA}~\cite{misc_ppg-dalia_495}: this dataset was recorded with a wrist-worn smartwatch Empatica E4 (provides wrist PPG signal) and a chest-worn device RespiBAN (offers ground truth data for HR and HRV derived from ECG) on 15 subjects. Therefore, we utilized it to evaluate HR and HRV estimation. Similarly to the PTT-PPG dataset, PPG-DaLiA was collected under various activities and we only used the data recorded during stationary periods for our evaluation. 

\textbf{WESAD}~\cite{schmidt2018introducing}: similar to PPG-DaLiA, the WESAD dataset also contains wrist PPG signals along with ground truth data for HR and HRV from ECG signals, and RR from respiration signals. The data was collected from 15 participants engaging in sedentary activities such as sitting, watching, and speaking. Thus, we utilized most of the data for the evaluation of HR estimation, HRV estimation, and RR estimation. 

\textbf{Graphene-HGCPT}~\cite{kireev2022continuous}: this dataset comprises a PPG signal recorded with a wrist-worn sensor alongside corresponding continuous arterial blood pressure (ABP) recordings obtained using the medical-grade Finapres NOVA device's finger cuff that estimates brachial BP. The data was collected from 7 young, healthy individuals, each underwent a series of lower motion activities to intentionally elevate blood pressure levels. We utilized this data to evaluate both Systolic BP (SBP) and Diastolic BP (DBP) estimation. 

\subsubsection{Heart Rate and Heart Rate Variability Estimation} 

We estimated HR in beats per minute (bpm) based on the number of detected pulse \textbf{valleys} (PPG) in the processed windows (8s), and used Mean Absolute Error (MAE, in bpm) to assess estimation performance. HRV metrics were calculated from the time differences between consecutive pulse \textbf{valleys}~\cite{han2022real, shin2009adaptive}. We selected two typical metrics, including Root Mean Square of the Successive Differences (RMSSD, in ms), and Standard Deviation of R-R Intervals (SDRR, in ms) to quantify HRV estimation accuracy.

From \Cref{tab:downstream-Results}, we can observe that across all datasets, the enhanced PPG signal consistently exhibits lower errors in all metrics for both HR (21.96\% on average) and HRV estimation  (45.70\% in RMSSD and 40.53\% in SDRR). This improvement can be attributed to \SysName's ability to compensate for temporal shifts induced by variations in contact pressure during the morphology refinement process. By refining the morphology, the enhanced signal provides more accurate heartbeat timing information, resulting in enhanced HR and HRV estimation accuracy. The self-collected data obtains lower performance yet larger improvement as it was collected under more diverse CP levels, while the CP in public dataset are relatively stable.

\subsubsection{Respiratory Rate Estimation}

We adopted a deep learning approach as introduced by~\cite{osathitporn2023rrwavenet}. Specifically, we employed five-fold cross-validation, ensuring that test folds contain different subjects and cover all subjects from the processed dataset. Both the original (distorted) and enhanced signals were segmented as $16s$ windows and then trained under the same model architecture. The results are reported as the average and standard deviation of the testing MAE in breaths per minute (brpm) across five folds, as shown in~\Cref{tab:downstream-Results}. Evidently, using enhanced signals notably outperformed the distorted signals by 6.85\%. The reason is that during the morphology transformation process, \SysName can regulate the amplitude shifts caused by changes in contact pressure. As a result, the rhythm information, which is crucial for RR estimation, is better preserved in the enhanced signal.

\subsubsection{Blood Pressure Estimation}

For BP estimation, we followed the commonly used machine learning-based approaches for SBP and DBP prediction (mmHg)~\cite{el2020review, 9558767, gonzalez2023benchmark}. In detail, we extracted multiple PPG waveform features such as SP, SW, and DW, along with other features such as HR and HRV, from the PPG windows (8s). These features were then utilized to train an Adaboost model for BP estimation. However, since many distorted PPG signals lack the diastolic peak, we could only extract a limited set of waveform features\footnote{The features include: SP, pulse area (PA), PA/SP, SW, DW, mean R-R intervals, SDRR, RMSSD, the percentage of successive R-R intervals differing by more than 50 ms (PNN50), and High Frequency (HF) power in spectral analysis.}. On the other hand, thanks to \SysName's ability to restore the diastolic peak and dicrotic notch, we were able to extract five additional waveform features: DN, NT, DP, DT, DT-SW. 

\begin{table*}
\centering
\caption{Physiological estimation performance with and without our model transformation across all datasets (MAE).}
\footnotesize
\setlength\tabcolsep{3pt}
\label{tab:downstream-Results}
\begin{tabular}{lllccccccc}
\toprule
\multirow{2}{*}{\textbf{Dataset}} & \multirow{2}{*}{\textbf{Size (min)}} & \multirow{2}{*}{\textbf{PPG Input}} & \multirow{2}{*}{\textbf{HR (bpm)}} & \multicolumn{2}{c}{\textbf{HRV}} & \multirow{2}{*}{\textbf{RR (brpm)}} & \multicolumn{2}{c}{\textbf{BP}} \\
\cmidrule(lr){5-6} 
\cmidrule(lr){8-9} 

& & & & \textbf{RMSSD (ms)} & \textbf{SDRR (ms)} & & \textbf{SBP (mmHg)} & \textbf{DBP (mmHg)} \\

\midrule
\multirow{2}{*}{Self-Collected} & \multirow{2}{*}{600} & Distorted & 1.84 $\pm$ 1.17 & 52.05 $\pm$ 40.89 & 31.03 $\pm$ 22.68 & -- & -- & --\\
& & \textbf{Enhanced} & \textbf{1.12 $\pm$ 1.00} & \textbf{9.43 $\pm$ 6.59} & \textbf{8.04 $\pm$ 4.78} & -- & -- & --\\
\midrule
\multirow{2}{*}{PTT-PPG} & \multirow{2}{*}{399} & Distorted & 0.54 $\pm$ 0.18 & 5.37 $\pm$ 1.70 & 3.58 $\pm$ 1.20 & -- & -- & --\\
& & \textbf{Enhanced} & \textbf{0.50 $\pm$ 0.12} & \textbf{4.97 $\pm$ 1.15} & \textbf{3.28 $\pm$ 1.04} & -- & -- & --\\
\midrule
\multirow{2}{*}{PPG-DaLiA} & \multirow{2}{*}{554} & Distorted & 1.01 $\pm$ 0.32 & 17.79 $\pm$ 6.18 & 13.87 $\pm$ 4.65 & -- & -- & --\\
& & \textbf{Enhanced} & \textbf{0.88 $\pm$ 0.28} & \textbf{11.20 $\pm$ 3.28} & \textbf{9.91 $\pm$ 3.13} & -- & -- & --\\
\midrule
\multirow{2}{*}{WESAD} & \multirow{2}{*}{2238} & Distorted & 1.09 $\pm$ 0.11 & 26.97 $\pm$ 12.21 & 17.79 $\pm$ 6.49 & 2.92 $\pm$ 0.43 & -- & --\\
& & \textbf{Enhanced} & \textbf{0.78 $\pm$ 0.09} & \textbf{12.29 $\pm$ 2.86} & \textbf{9.05 $\pm$ 1.90} & \textbf{2.72 $\pm$ 0.35} & -- & --\\
\midrule
\multirow{2}{*}{Graphene-HGCPT} & \multirow{2}{*}{753} & Distorted & -- & -- & -- & -- & 7.00 $\pm$ 0.26 & 4.63 $\pm$ 0.14\\
& & \textbf{Enhanced} & -- & -- & -- & -- & \textbf{6.65 $\pm$ 0.20} & \textbf{4.46 $\pm$ 0.15}\\
\bottomrule
\end{tabular}
\vspace{-0.3cm}
\end{table*}

\Cref{tab:downstream-Results} displays testing MAE average and standard deviation for SBP and DBP across five folds. We can observe that the enhanced PPG signals with refined typical features and additional diastolic features provide better predictions (5\% in SBP, and 3.67\% in DBP) of blood pressure compared to using the distorted PPG (only typical features). This improvement can be attributed to the enhanced PPG signals having more accurate physiological characteristics, which facilitate more accurate feature extraction used for training the model. 

\textbf{Remarks: }Although \SysName was trained with the self-collected data, its superior performance on public datasets with various physiological tasks demonstrates its strong generalizability\footnote{Compared to our self-collected dataset, the relatively smaller performance increase on public datasets is because we only select the stationary segments that exhibit limited variations in CP (and consequently, on signal morphology). In other words, they are less affected by CP.}. Additionally, \SysName can be used as a \textbf{plug-in API} to refine any contact pressure-distorted PPG signal. For example, it could be integrated into smartwatches as a pre-processing step to improve PPG signal quality before HR/HRV estimation, resulting in enhanced performance.  

\subsection{Performance in In-the-wild Settings} \label{itw}

\subsubsection{Experiment Setup}

The above evaluations were conducted under controlled conditions and may not fully represent real-world scenarios. Specifically, in the self-collected data, we manually adjusted the contact pressure of the wrist PPG sensor using a clamp device, which does not accurately reflect the \textit{natural} variations in contact pressure caused by posture changes in real-life situations. Additionally, public datasets were typically collected in confined environments, where subjects are instructed to perform specific activities or maintain certain states~\cite{mehrgardt2022pulse, misc_ppg-dalia_495,schmidt2018introducing, kireev2022continuous}.

\begin{figure}
\centering
\begin{minipage}[h]{0.38\linewidth}
    \centering
    \includegraphics[width=0.9\linewidth, angle=0]{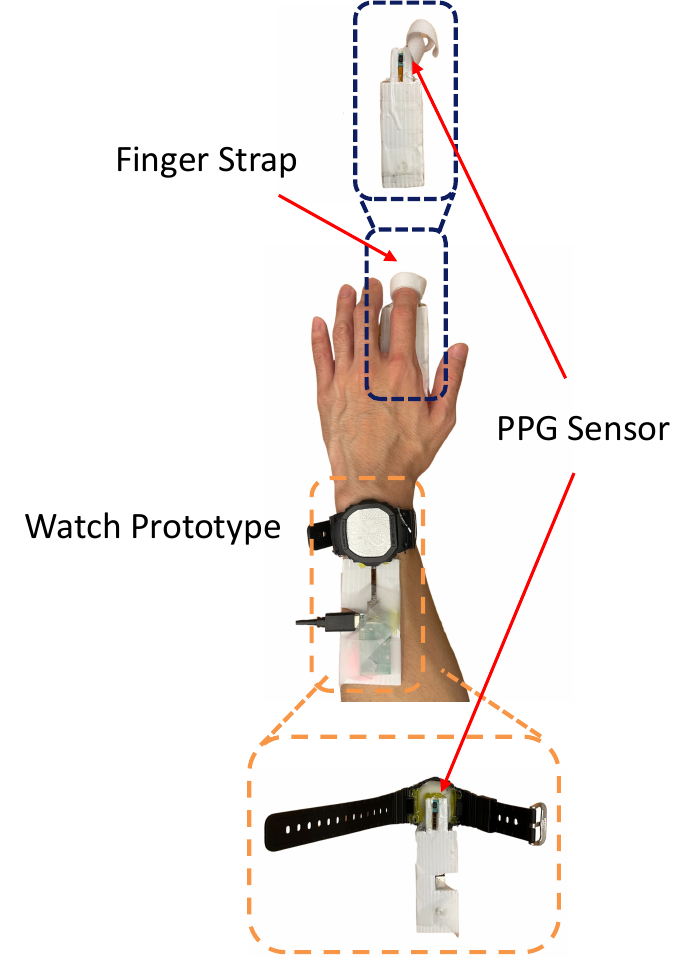}
    \caption{Illustration of the upgraded watch prototype for in-the-wild study.}
    \label{fig:itw_prototype}
    \vspace{-0.2cm}
\end{minipage}%
\hfill
\begin{minipage}[h]{0.58\linewidth}
    \centering
    \vspace{-0.3cm}
    \captionof{table}{HR/HRV performance on in-the-wild study (MAE).}
    \vspace{-0.3cm}
    \footnotesize
    \setlength\tabcolsep{6.5pt}
    \renewcommand{\arraystretch}{1.1}
    \label{tab:itw_results}
    \begin{tabular}{llccc}
    \toprule
     & \multirow{2}{*}{\textbf{PPG Input}} & \multirow{2}{*}{\textbf{HR (bpm)}} & \multicolumn{2}{c}{\textbf{HRV}} \\
    \cmidrule(lr){4-5}
    & & & \textbf{RMSSD (ms)} & \textbf{SDRR (ms)} \\
    \midrule
    \multirow{3}{*}{Subject 1} & Distorted & 1.47 $\pm$ 0.31 & 15.10 $\pm$ 3.24 & 12.44 $\pm$ 3.80 \\
    & \textbf{Enhanced} & \textbf{0.52 $\pm$ 0.11} & \textbf{5.21 $\pm$ 1.13} & \textbf{6.65 $\pm$ 2.90} \\
    & Reference & 0.51 $\pm$ 0.10 & 5.13 $\pm$ 1.04 & 6.26 $\pm$ 2.44 \\
    \midrule
    \multirow{3}{*}{Subject 2} & Distorted & 1.09 $\pm$ 0.41 & 11.14 $\pm$ 4.26 & 11.03 $\pm$ 3.50 \\
    & \textbf{Enhanced} & \textbf{0.87 $\pm$ 0.26} & \textbf{8.82 $\pm$ 2.70} & \textbf{10.18 $\pm$ 2.88} \\
    & Reference & 0.94 $\pm$ 0.33 & 9.58 $\pm$ 3.33 & 10.64 $\pm$ 3.11 \\
    \midrule
    \multirow{3}{*}{Subject 3} & Distorted & 1.55 $\pm$ 0.58 & 16.08 $\pm$ 5.97 & 18.12 $\pm$ 5.51 \\
    & \textbf{Enhanced} & \textbf{1.26 $\pm$ 0.62} & \textbf{13.18 $\pm$ 6.42} & \textbf{17.13 $\pm$ 5.66} \\
    & Reference & 1.29 $\pm$ 0.64 & 13.45 $\pm$ 6.66 & 17.03 $\pm$ 5.65 \\
    \midrule
    \multirow{3}{*}{Subject 4} & Distorted & 0.57 $\pm$ 0.23 & 5.89 $\pm$ 2.44 & 6.79 $\pm$ 1.33 \\
    & \textbf{Enhanced} & \textbf{0.36 $\pm$ 0.06} & \textbf{3.70 $\pm$ 0.57} & \textbf{6.33 $\pm$ 0.92} \\
    & Reference & 0.32 $\pm$ 0.07 & 3.30 $\pm$ 0.68 & 6.21 $\pm$ 0.64 \\
    \midrule
    \multirow{3}{*}{Subject 5} & Distorted & 1.57 $\pm$ 0.58 & 16.00 $\pm$ 6.09 & 15.43 $\pm$ 2.75 \\
    & \textbf{Enhanced} & \textbf{0.89 $\pm$ 0.16} & \textbf{8.95 $\pm$ 1.59} & \textbf{12.28 $\pm$ 2.78} \\
    & Reference & 0.86 $\pm$ 0.20 & 8.51 $\pm$ 1.96 & 11.42 $\pm$ 3.04 \\
    \bottomrule
    \end{tabular}
     \vspace{-0.5cm}
\end{minipage}
\end{figure}

To address these issues, we conducted a longitudinal in-the-wild study with five participants over five consecutive days. First, to ensure user comfort during the experiments, we upgraded our prototype into a smartwatch form factor, as shown in \Cref{fig:itw_prototype}. Specifically, we embedded the wrist PPG sensor at the bottom of an empty watch case, allowing participants to wear it with an adjustable strap for their preferred tightness. The finger PPG sensor was secured with a clip, and participants also wore an ECG chest band to obtain ground truth data for HR and HRV.

Each participant wore the devices for 30 minutes daily, during which they were free to participate in any desk-related sedentary activities such as working on a computer, reading a book, watching videos on the phone, and sitting on a chair. These activities naturally led to posture changes, causing varying degrees of morphology distortion in the wrist PPG signal. After collecting the data, we applied the evaluation pipeline for downstream tasks to the wrist PPG signal to compare the HR and HRV estimation performance. Meanwhile, we also calculated the HR and HRV results for the finger PPG signal (denoted as Reference), which serves as a reference performance achievable using finger-based PPG devices (e.g. pulse oximeter).

\subsubsection{Performance on HR and HRV Estimation}~\Cref{tab:itw_results} compares the HR and HRV estimation performance (MAE) among the distorted wrist PPG signals, enhanced wrist PPG signals with \SysName, and the finger PPG signals (reference) from our in-the-wild data. The values are averaged over all the data collected during the five days. We can observe that: 1) for all the HR and HRV metrics, the enhanced wrist PPG signal significantly outperforms the distorted signal, and the performance is very close to that of the finger PPG signal. This is because the data collected with the developed clamp prototype under various CPs effectively captures comprehensive variations of the PPG morphology to operate in realistic conditions; 2) the results are consistent across different subjects, indicating the robustness and generalizability of our approach; 3) an interesting observation is that the enhanced signal can even outperform the reference signal in some cases (e.g., Subject 2). The reason is that the reference signal is from a finger PPG, which can still suffer from poorer contact pressure depending on the behaviors of the users despite an initial perfect fit, leading to degraded performance.

\subsection{System Performance}

Since \SysName can be used as a plug-in API, we investigate its system performance regarding peak memory usage, latency, and energy consumption. The measurements were carried out on three devices: a MacBook Air M1, a Raspberry Pi 5, and a Xiaomi 13 smartphone.~\Cref{tab:system_performance} presents the results averaged over 100 inferences. We can observe that \SysName operates efficiently, with peak memory usage remaining under 180~MB across three devices, which translates to only 2\% of the total memory budget even on the Raspberry Pi 5. The Macbook Air exhibited the lowest memory usage, likely due to its optimized memory management and superior hardware architecture. In terms of latency, \SysName processes an eight-second PPG sample in under 100 ms on all platforms, indicating its capability for real-time inference. Additionally, \SysName is energy-efficient; for instance, on the Xiaomi 13 smartphone, it consumes only 0.11~J, a negligible fraction of the overall battery capacity (71,280~J). These results demonstrate that \SysName is a lightweight model, making it well-suited for deployment on typical edge-scale devices.   

\begin{table}[h]
\centering
\small
\setlength\tabcolsep{8pt}
\vspace{-0.3cm}
\caption{System performance of \SysName.}
\vspace{-0.3cm}
\label{tab:system_performance}
\begin{tabular}{lccc}
\toprule
& \textbf{Memory} & \textbf{Latency} & \textbf{Energy } \\
& \textbf{(MB)} & \textbf{(ms)} & \textbf{ (J/sample)} \\
\midrule
MacBook Air M1 & 85.85 & 11.65 & --  \\
Raspberry Pi 5 & 174.36  & 69.74 & 0.29 \\
Xiaomi 13 Phone & 127.30  & 60.12  & 0.11 \\ 
\bottomrule
\vspace{-0.6cm}
\end{tabular}
\end{table}

%% file: sections/08-Discussion.tex
\section{Discussion}  \label{discussion}
This section elaborates on the design considerations in the study and outlines potential avenues for future exploration.

\textbf{Demographic considerations}. In this work, we collected data from young, healthy participants who generally present the ideal PPG morphology. However, even with optimal contact pressure, the PPG morphology of older individuals (due to vascular aging~\cite{Millasseau2002DeterminationOA}) or those with certain cardiovascular diseases~\cite{Angius2012MyocardialIA}) (e.g., arrhythmias) deviates from this ideal morphology. Consequently, the trained model in this work may not be directly applicable to these populations. However, the proposed framework remains valid. For these demographic groups, future researchers can collect appropriate data following our experimental conditions (the only difference is that the finger PPG signal under optimal contact pressure may not present the ideal morphology) and fine-tune the model. 

\textbf{Performance improvement with multiple PPG channels}. PPG sensors can operate in different color channels including green, red, and infrared, with each color penetrating to different depths and measuring different types of blood vessels~\cite{Liu2016PPGColorDepth}. We focus on transforming PPG signals from the green light channel, as it is typically used in smartwatch settings. However, the proposed framework can also be applied to other color channels, where PPG sensors might be deployed in different measurement sites (e.g., ear). Moreover, given that different colors exhibit distinct optical characteristics, it is possible to leverage the inherent correlation of PPG waveforms across different color channels to enhance the performance and reliability of waveform transformation. For instance, if the PPG waveform from the red channel is corrupted due to real-world factors, the other channels can still facilitate the transformation process.

\textbf{Potential of using contact pressure for motion artifact removal}. This work focuses on mitigating the impact of contact pressure on PPG waveforms under sedentary conditions. In scenarios involving intensive motion, sensor displacements or user movements also induce changes in contact pressure. Therefore, it is promising to incorporate contact pressure measurements into the process of PPG motion artifact removal. However, the major difference with this work is that in \SysName, contact pressure remains stable within a heartbeat cycle (i.e., a complete PPG cycle), while it varies within the PPG cycle in motion cases. The superimposition of variations due to internal blood volume and external contact pressure makes the problem extremely challenging and is left as future work.

%% file: sections/07-RelatedWorks.tex
\section{Related Work} \label{relatedworks}

\subsection{Impact of Contact Pressure }
Recent works have begun to analyze the impact of contact pressure between human skin and the sensor on the PPG signal~\cite{Chandrasekhar2020ContactPressure, SIRKIA2024PPGandCP, Pi2021PressureOnWaveforms}. Specifically, as human skin is soft, varying contact pressures cause different degrees of tissue deformation and thus varying lengths of light propagation. Consequently, the signal quality of the PPG signal is affected. For example, May et al. experimentally demonstrated the correlation between contact pressure and the amplitudes of AC and DC components, indicating the existence of an optimal contact pressure that generates PPG signals with the highest signal-to-noise ratio (i.e., ideal morphology)~\cite{May2021InVitroVessel}. In addition, Scardulla et al. investigated the impact of contact pressure on the HR estimation accuracy at various exercise intensities using a smartwatch-like device, finding that higher contact pressures enable more accurate heart rate estimation~\cite{Scardulla2020LoadCellHR}. However, these works merely highlight the issue of contact pressure without connecting it to real-world scenarios and proposing solutions to address it. Our work delved deeper into the problem of contact pressure and proposed a generic approach that can convert PPG signals measured at sub-optimal skin-sensor contact into ideal waveforms, and demonstrated its effectiveness with comprehensive experiments.

\subsection{Combating Motion Artifacts }
Motion artifacts are unavoidable in wearables like smartwatches, prompting numerous efforts to mitigate their effects. Conventional signal processing techniques have been used to address motion-related distortions in PPG signals. For example, adaptive filtering methods, such as the adaptive step-size least mean squares filter~\cite{Ram2012_2}, utilize motion data from an accelerometer as a reference~\cite{Pollreisz2019AccelerometerCompensation} to remove these artifacts. Recently, deep learning techniques, including convolutional neural networks~\cite{Reiss2019,Mehrgardt2021PTTPPG}, long and short-term memory~\cite{Biswas2019}, generative adversarial-based networks~\cite{wang2022ppg, Zheng2022_6, afandizadeh2023accurate, mahmud2024wearable, long2023reconstruction}, have been employed for motion-distorted PPG signals, achieving downstream results that surpass traditional signal processing approaches. 

Our work identifies PPG morphology variations caused by contact pressure changes during sedentary activities, an aspect often overlooked in previous research. Unlike prior waveform reconstruction efforts that primarily focus on recovering the systolic peak for HR estimation, our study emphasizes the reconstruction of finer-grained PPG waveform features. This approach aims to achieve an ideal PPG morphology, enabling more advanced downstream applications that require sophisticated features.

%% file: sections/09-Conclusion.tex
\section{Conclusion} \label{conclusion}

In this study, we identified the PPG morphology distortions due to contact pressure changes during real-life sedentary activities and proposed an effective approach to address the issue. Specifically, we introduced \SysName, a novel framework designed to transform low-quality PPG waveforms measured under suboptimal contact pressure into ideal ones. \SysName comprises data collection, a suite of well-crafted signal processing techniques, and a lightweight deep-learning model with specialized designs. The extensive results demonstrated that \SysName not only enhances PPG signal quality but also improves sensing performance across four typical PPG downstream tasks. Notably, the performance on public PPG datasets and in-the-wild experiment highlights \SysName's versatility in transforming distorted PPG signals, showcasing its significance in achieving accurate and robust physiological sensing under realistic conditions.